\documentclass{amsart}
\usepackage{amsmath,amsthm,amssymb,amscd,cite}

\theoremstyle{definition}

\theoremstyle{remark}

\numberwithin{equation}{section}

\newtheorem{thm}{Theorem}[section]

\newtheorem{lem}[thm]{Lemma}

\newtheorem{rmk}[thm]{Remark}

\begin{document}

\title{Dilation, Discrimination and Uhlmann's Theorem of Link Products of Quantum Channels}

\author{Qiang Lei}
\address{Qiang Lei. School of Mathematics Sciences, Harbin Institute of Technology, Harbin 150001, China.}
\curraddr{}
\email{leiqiang@hit.edu.cn}


\author{Liuheng Cao}
\address{Liuheng Cao. School of Mathematics Sciences, Harbin Institute of Technology, Harbin 150001, China.} \email{2826142155@qq.com}

\author{Asutosh Kumar}
\address{Asutosh Kumar. Department of Physics, Gaya College, Magadh University, Rampur, Gaya 823001, India}
\address{Asutosh Kumar. Vaidic and Modern Physics Research Centre, Bhagal Bhim, Bhinmal, Jalore 343029, India}\email{asutoshk.phys@gmail.com}

\author{Junde Wu}
\address{Junde Wu. School of Mathematics Sciences, Zhejiang University, Hangzhou 310027, China.}
\email{wjd@zju.edu.cn}

\curraddr{}
\email{}



\begin{abstract}
The study of quantum channels is the most fundamental theoretical problem in quantum information and quantum communication theory. The link product theory of quantum channels is an important tool for studying quantum networks. In this paper, we establish the Stinespring dilation theorem of the link product of quantum channels in two different ways, discuss the discrimination of quantum channels and show that the distinguishability can be improved by self-linking each quantum channel $n$ times as $n$ grows. We also find that the maximum value of Uhlmann's theorem can be achieved for diagonal channels.
\end{abstract}
\maketitle

\section{Introduction}
\label{Introduction}
In quantum computation and information processing, a number of measurements and transformations are called into action depending on the specific task to be performed. A quantum network, which is an assembly or combination of a sufficient number of elementary circuits that transform or measure quantum states, can be used in a number of different ways to accomplish different tasks. Various transformations are brought about by quantum operations or quantum channels. A quantum channel is a linear map or quantum operation that is completely positive and trace-preserving. It can be imagined as a unitary interaction of a system with an environment. Quantum channels essentially describe how quantum information is transformed when it passes through physical systems, including potential sources of noise and errors.

Quantum channel dilation refers to the process of constructing a new and better quantum channel by introducing additional qubits on a noisy quantum channel. Quantum channel dilation is widely used in quantum key distribution, quantum state transmission, and quantum error correction. It can facilitate realization of efficient quantum communication and information processing \cite{stinespring, continuity,fock, strong}. According to the Stinespring dilation theorem, any quantum channel can be implemented as an isometric channel in a larger system.

In order to ensure the reliability and security of information transmission,  it is necessary to distinguish different types of quantum channels in quantum communication.  This problem has been considered by many people \cite{fidelity,memory, base,conditions,channel,coherent,quantum, conclusive}. Quantum channel discrimination technology can be applied to various quantum communication protocols, such as quantum key distribution, quantum remote state preparation, and quantum algorithms.

The link product theory of quantum channels \cite{fram} is an important tool for studying quantum networks \cite{fram,cloning,circuit,optimal,network,Xiao}. Link product is the operation that connects two quantum networks. In this paper, we establish the Stinespring dilation theorem of link product of quantum channels in two different ways. We discuss the discrimination of quantum channels and show that the distinguishability can be improved by considering the link product of each quantum channel $n$ times for large $n$. We also find that the maximum value of Uhlmann's theorem can be achieved for diagonal channels.

\section{Stinespring dilation theorem of link product of quantum channels}

\subsection{Preliminaries and Notation}

Let $\mathcal{H}$ be a finite dimensional Hilbert space and $\mathcal{L}(\mathcal{H})$ be the set of linear operators on $\mathcal{H}$. $\mathcal{L}(\mathcal{H}_0, \mathcal{H}_1)$ denotes the set of linear operators from $\mathcal{H}_0$ to $\mathcal{H}_1$. The set of linear maps from $\mathcal{L}(\mathcal{H}_0)$ to $\mathcal{L}(\mathcal{H}_1)$ is denoted by $\mathcal{L}(\mathcal{L}(\mathcal{H}_0), \mathcal{L}(\mathcal{H}_1))$. $I_0 \equiv I_{\mathcal{H}_0}$ is the identity operator in $\mathcal{H}_0$, 
$\mathcal{I}_0 \equiv \mathcal{I}_{\mathcal{L}(\mathcal{H}_0)}$ is the identity map on $\mathcal{L}(\mathcal{H}_0)$, $\textup{Tr}_{0} \equiv \textup{Tr}_{\mathcal{H}_0}$ is the partial trace over $\mathcal{H}_0$, and so on. Furthermore, $M_{xy}^T$ denotes the transpose over $\mathcal{H}_x \otimes \mathcal{H}_y$ and $M_{xy}^{T_y}$ denotes the partial transpose over $\mathcal{H}_y$.

Choi isomorphism is a one-to-one correspondence map $\mathfrak{C}:\ \mathcal{L}(\mathcal{L}(\mathcal{H}_0),
\mathcal{L}(\mathcal{H}_1)) \rightarrow \mathcal{L}(\mathcal{H}_1 \otimes \mathcal{H}_0)$ defined as
\begin{align}\label{5}
\mathfrak{C}:\ \mathcal{M} \mapsto M_{10} := (\mathcal{M} \otimes \mathcal{I}_0)(|I_0\rangle\rangle \langle\langle I_0|),
\end{align}
where $\mathcal{I}_0$ is the identity map on $\mathcal{L}(\mathcal{H}_0), |I_0\rangle\rangle \langle\langle I_0| = \sum_{n,m}|n\rangle|m\rangle\langle n|\langle m|$, and
$\{|m\rangle\}$ is a fixed orthonormal basis in $\mathcal{H}_0$.
$M_{10}$ is called the Choi operator of $\mathcal{M}$. 

Moreover, the inverse map $\mathfrak{C}^{-1} : \mathcal{L}(\mathcal{H}_1 \otimes \mathcal{H}_0) \longrightarrow \mathcal{L}(\mathcal{L}(\mathcal{H}_0),\mathcal{L}(\mathcal{H}_1)) $ is decided by
\begin{equation}\label{6}
\mathcal{M}(X)=[\mathfrak{C}^{-1}(M_{10})](X) =\textup{Tr}_{0}[(I_1 \otimes X^T)M_{10}],
\end{equation}
where $ M_{10} \in \mathcal{L}(\mathcal{H}_1 \otimes \mathcal{H}_0)$ and $X \in \mathcal{L}(\mathcal{H}_0)$.

If a completely positive map $\mathcal{M}\in\mathcal{L}((\mathcal{L}(\mathcal{H}_0),\mathcal{L}(\mathcal{H}_1))$ is trace-preserving, i.e., $\textup{Tr}(\mathcal{M}(X))=\textup{Tr}(X)$, then it is called a quantum channel. When $\mathcal{M}\in\mathcal{L}((\mathcal{L}(\mathcal{H}_0),\mathcal{L}(\mathcal{H}_1))$ is a quantum channel, we have $\textup{Tr}_1[M_{10}]=I_0$. On the other hand, there exists $\{A_i|A_i\in\mathcal{L}(\mathcal{H}_0,\mathcal{H}_1)\}$ with $\sum_i A_i^{\dagger}A_i =I_0 $ such that $\mathcal{M}(X)=\sum_i A_iXA_i^{\dagger}$.

A relevant class of channels are the isometric channels $\mathcal{V}:\mathcal{L}(\mathcal{L}(\mathcal{H}_0),\mathcal{L}(\mathcal{H}_1))$, defined as
\begin{equation}
\mathcal{V}(\rho):=V\rho V^{\dagger}\
\end{equation}
where $V\in \mathcal{L}(\mathcal{H}_0,\mathcal{H}_1)\ and\  V^{\dagger}V=I_0$ \cite{book1}.

\subsection{Stinespring dilation theorem of composition of quantum channels}

The Stinespring dilation theorem states that every quantum channel can be realized as an isometric channel on a larger system.
\begin{lem} (Stinespring dilation theorem)\cite{realization}
Let $M_{10}\in\mathcal{L}(\mathcal{H}_1\otimes\mathcal{H}_0)$ be the Choi operator associated with the quantum channel $\mathcal{M}\in\mathcal{L}(\mathcal{L}(\mathcal{H}_0),\mathcal{L}(\mathcal{H}_1))$, and let $\mathcal{H}_A$ be the Hilbert space $\mathcal{H}_A=\textup{Supp}(M_{10})$. Then, a minimal dilation $V:\ \mathcal{H}_0 \longrightarrow \mathcal{H}_1 \otimes \mathcal{H}_{\mathcal{A}}$ is given by
	\begin{equation}
	V=(I_1 \otimes (M_{10}^T)^{1/2})(|I_1\rangle\rangle \otimes I_0)
	\end{equation}
	and
	\begin{equation}
	\mathcal{M}(\rho)=\textup{Tr}_A[\mathcal{V}(\rho)]=\textup{Tr}_A[V\rho V^{\dagger}]
	\end{equation}
	where $V$ is called the Stinespring dilation of $\mathcal{M}$.
\end{lem}

Now we establish the Stinespring dilation of composition of quantum channels, which depends on the Stinespring dilation $V_1$ of $\mathcal{M}$ and $V_2$ of $\mathcal{N}$.
\begin{thm}
	Let $\mathcal{M}\in \mathcal{L}(\mathcal{L}(\mathcal{H}_0),\mathcal{L}(\mathcal{H}_1))$ and $\mathcal{N}\in\mathcal{L}(\mathcal{L}(\mathcal{H}_1),\mathcal{L}(\mathcal{H}_2))$ be two quantum channels and $ M_{10}\in\mathcal{L}(\mathcal{H}_1 \otimes \mathcal{H}_0),N_{21}\in \mathcal{L}(\mathcal{H}_2 \otimes \mathcal{H}_1)$ be their corresponding Choi operators.
	Then there exists an ancillary Hilbert space $\mathcal{H}_A$ and an isometry $V:\mathcal{H}_0 \longrightarrow \mathcal{H}_2 \otimes \mathcal{H}_{A},V^{\dagger} V=I_0$ such that
	\begin{equation}
	\mathcal{N}\circ\mathcal{M}(\rho)=\textup{Tr}_{A}[\mathcal{V}(\rho)]=\textup{Tr}_{A}[V \rho V^{\dagger}]
	\end{equation}
	for any $\rho\in\mathcal{L}(\mathcal{H}_0)$.
\end{thm}
\begin{proof}
	By Lemma 2.1, there are isometry dilations  $V_1:\mathcal{H}_0 \longrightarrow \mathcal{H}_1 \otimes \mathcal{H}_{A_1}$ and $V_2:\mathcal{H}_1  \longrightarrow \mathcal{H}_2 \otimes \mathcal{H}_{A_2}$,
	\begin{equation}
	V_1=(I_1 \otimes (M_{10}^T)^{1/2})(|I_1\rangle\rangle \otimes I_0),
	\end{equation}
	\begin{equation}
	V_2=(I_2 \otimes (N_{21}^T)^{1/2})(|I_{2}\rangle\rangle \otimes I_{1}),
	\end{equation}
	where $\mathcal{H}_{A_1}=\textup{Supp}(M_{10})$ and $\mathcal{H}_{A_2}=\textup{Supp}(N_{21})$. Furthermore, let $\mathcal{H}_A=\mathcal{H}_{A_1}\otimes \mathcal{H}_{A_2}$.
	
	Define $V:\mathcal{H}_0 \longrightarrow \mathcal{H}_2 \otimes \mathcal{H}_{A_1} \otimes \mathcal{H}_{A_2}$ as $$\ V:=(V_2\otimes I_{A_1}) \circ V_1$$ where $I_{A_1}$ is the identity operator on $\mathcal{H}_{A_1}$. Since $M_{10} \geq 0$, we have $M_{10}^{\dagger}=M_{10}$, that is, $M_{10}^T = M_{10}^{\ast}$. Similarly, $N_{21}^T = N_{21}^{\ast}$. Then we have
	\begin{align}
	V^{\dagger}V =(V_2\circ V_1)^{\dagger}(V_2\circ V_1)=V_1^{\dagger}V_2^{\dagger}V_2 V_1=I_0,
	\end{align}
	and
	\begin{align*}
	\textup{Tr}_{A}[V \rho V^{\dagger}] =&\textup{Tr}_{A}[(I_2 \otimes (N_{21}^{\textup{T}})^{1/2})(|I_2\rangle\rangle \otimes I_1)(I_1 \otimes (M_{10}^{\textup{T}})^{1/2})(|I_1\rangle\rangle \otimes I_0)\rho(\langle\langle I_1 \otimes I_0) \\
	&(I_1 \otimes (M_{10}^{\ast})^{1/2})(\langle\langle I_2|\otimes I_1)(I_2 \otimes (N_{21}^{\ast})^{1/2})]\\
	=&\textup{Tr}_{0,1,2}[(I_2 \otimes (N_{21}^{\textup{T}})^{1/2})(I_1 \otimes (M_{10}^{\textup{T}})^{1/2})(|I_1\rangle\rangle  \langle\langle I_1 \otimes \rho)(I_1 \otimes (M_{10}^{\ast})^{1/2})\\
	&(|I_2\rangle\rangle\langle\langle I_2| \otimes I_1)(I_2 \otimes (N_{21}^{\ast})^{1/2})]\\
	=&\textup{Tr}_{0,1,2}[(I_2 \otimes N_{21}^{\textup{T}})(I_1 \otimes (M_{10}^{\textup{T}})^{1/2})(|I_1\rangle\rangle  \langle\langle I_1 \otimes \rho)(I_1 \otimes (M_{10}^{\ast})^{1/2})\\
	&(|I_2\rangle\rangle\langle\langle I_2| \otimes I_1)]\\
	=&\textup{Tr}_{1,2}[(I_2 \otimes (N_{21}^{\textup{T}}) \mathcal{M}(\rho)^{\textup{T}})\times |I_1\rangle\rangle\langle\langle I_1| \times (|I_2\rangle\rangle\langle\langle I_2| \otimes I_1)]\\
	=&\textup{Tr}_2[I_2\otimes \textup{Tr}_1[N_{21}^{\textup{T}}(I_2\otimes \mathcal{M}(\rho))] \times |I_2\rangle\rangle\langle\langle I_2|]\\
	=&\textup{Tr}_2[I_2\otimes(\mathcal{N}\circ\mathcal{M}(\rho))^{\textup{T}}]\times |I_2\rangle\rangle\langle\langle I_2|]\\
	=&\mathcal{N} \circ\mathcal{M}(\rho)
	\end{align*}
	where $\textup{Tr}_{0,1,2}$ denotes the partial trace over the Hilbert spaces $\mathcal{H}_0,\mathcal{H}_1$ and $\mathcal{H}_2$, respectively.
\end{proof}

This construction of  the Stinespring dilation of composition of quantum channels is simple to manipulate. However, the ancilla dimension is not minimum. Below we give another dilation of $\mathcal{N}\circ\mathcal{M}$ such that the ancillary space has the minimum dimension.

\begin{thm}
	Let $\mathcal{M}\in \mathcal{L}(\mathcal{L}(\mathcal{H}_0),\mathcal{L}(\mathcal{H}_1))$ and $\mathcal{N}\in\mathcal{L}(\mathcal{L}(\mathcal{H}_1),\mathcal{L}(\mathcal{H}_2))$ be two quantum channels and $ M_{10}\in\mathcal{L}(\mathcal{H}_1 \otimes \mathcal{H}_0),N_{21} \in \mathcal{L}(\mathcal{H}_2 \otimes \mathcal{H}_1)$ be their corresponding Choi operators. Let $\mathcal{H}_A =\textup{Supp}(\mathfrak{C}(\mathcal{N}\circ\mathcal{M}))$. Then there is a minimal isometry dilation $V:\mathcal{H}_0 \longrightarrow  \mathcal{H}_2 \otimes \mathcal{H}_{\mathcal{A}}$ is given by
	\begin{align}\label{16}
	V=\textup{Tr}_1[(N_{21}^{\tfrac{1}{2}T} \otimes I_0)(I_2 \otimes M_{10}^{\tfrac{1}{2}T_0})(|I_2\rangle\rangle \otimes I_1 \otimes I_0)]
	\end{align}
	such that $$\mathcal{N}\circ\mathcal{M}(\rho)=\textup{Tr}_A[V\rho V^{\dagger}].$$
\end{thm}
\begin{proof}
	\begin{align}
	\begin{split}
	V^{\dagger}V &=\textup{Tr}_1[(\langle\langle I_2| \otimes I_1 \otimes I_0)(I_2 \otimes M_{10}^{\tfrac{1}{2} T_0})
	(N_{21}^{\tfrac{1}{2}T} \otimes I_0)
	(N_{21}^{\tfrac{1}{2}T} \otimes I_0)(I_2 \otimes M_{10}^{\tfrac{1}{2}T_0})(|I_2\rangle\rangle \otimes I_1 \otimes I_0)] \\
	&=\textup{Tr}_{1,2}[(I_2 \otimes M_{10}^{\tfrac{1}{2} T_0})(N_{21}^T \otimes I_0)(I_2 \otimes M_{10}^{\tfrac{1}{2} T_0})]  \\
	&=\textup{Tr}_{1.2}[(N_{21}^T \otimes I_0)(I_2 \otimes M_{10}^{T_0})]  \\
	&=\textup{Tr}_2[(\mathfrak{C}(\mathcal{N}\circ\mathcal{M}))^T]  \\
	&=I_0
	\end{split}
	\end{align}
	and
	\begin{align}
	\begin{split}
	\textup{Tr}_{A}[V \rho V^{\dagger}] =&\textup{Tr}_{0,1,2}[(N_{21}^{\tfrac{1}{2}T} \otimes I_0)(I_2 \otimes M_{10}^{\tfrac{1}{2}T_0})(|I_2\rangle\rangle \otimes I_1 \otimes I_0)\rho(\langle\langle I_2| \otimes I_1 \otimes I_0)(I_2 \otimes M_{10}^{\tfrac{1}{2} T_0})  \\
	&(N_{21}^{\tfrac{1}{2}T} \otimes I_0)]  \\
	=&\textup{Tr}_{0,2}[\textup{Tr}_1[(N_{21}^T \otimes I_0)(I_2 \otimes M_{10}^{T_0})](|I_2\rangle\rangle \langle\langle I_2| \otimes \rho)]  \\
	=&\textup{Tr}_{0,2}[(I_2 \otimes \mathfrak{C}(\mathcal{N}\circ\mathcal{M})^T)(|I_2\rangle\rangle\langle\langle I_2| \otimes \rho)]  \\
	=&\textup{Tr}_2[(I_2 \otimes \textup{Tr}_0[(I_2\otimes\rho)(\mathfrak{C}(\mathcal{N}\circ\mathcal{M}))^T])\times(|I_2\rangle\rangle\langle\langle I_2|)]\\
	=&\textup{Tr}_2[(I_2\otimes(\mathcal{N}\circ\mathcal{M}(\rho))^T)|I_2\rangle\rangle\langle\langle I_2|]\\
	=&\mathcal{N}\circ\mathcal{M}(\rho).
	\end{split}
	\end{align}
\end{proof}

\begin{rmk}
The above two approaches have their own merits and demerits.
The dilation method in Theorem 2.2 corresponds to indirect composition dilation and is easier to implement. But it may cause waste of resource. The way in Theorem 2.3 is a direct composition dilation. Similar to the proof in \cite{network}, the isometric dilation given in Theorem 2.3 has a minimum ancilla dimension.  It can minimize resource waste but requires more computational and communication resources. For example, suppose $\textbf{rank}(\mathcal{M})=\textbf{rank}(\mathcal{N})=16$. In the case of indirect composition dilation, $\textbf{dim}(\mathcal{H}_{A_1})=16,\textbf{dim}(\mathcal{H}_{A_2})=16$. Although the required space resource $\mathcal{H}_A$ has a dimension of $256$, it belongs to two systems and only 16-dimensional calculations are required. On the other hand, in the case of direct composition dilation, $\textbf{dim}(\mathcal{H}_A)=\textbf{rank}(\mathcal{N}\circ\mathcal{M})\leq 256$. The dilated space belongs to only one system, and the computational dimension can reach up to 256. Therefore, in specific applications, we need to choose the appropriate dilation method according to the actual situation to achieve the optimal effect.
\end{rmk}

\subsection{Stinespring dilation theorem of link product of quantum channels}

If we consider quantum channels such that their input and output spaces are the tensor product of Hilbert spaces, it is possible to compose these channels only through inserting or pasting some of these spaces or identity channels. At this time, the composition of two quantum channels  $\mathcal{M}$ and $\mathcal{N}$  is denoted by  $\mathcal{N} \star \mathcal{M}$, and is said to be the link product of $\mathcal{M}$ and $\mathcal{N}$; it is an important tool of studying the quantum network theory \cite{network}.

It then follows that using Theorem 2.2 and Theorem 2.3, we can get the following Stinespring dilation theorems of link product of two quantum channels.

\begin{thm}
	Let $\mathcal{M} \in \mathcal{L}(\mathcal{L}(\mathcal{H}_0 \otimes \mathcal{H}_2),\mathcal{L}(\mathcal{H}_1 \otimes \mathcal{H}_3))$ and $\mathcal{N} \in \mathcal{L}(\mathcal{L}(\mathcal{H}_3 \otimes \mathcal{H}_5),\mathcal{L}(\mathcal{H}_4 \otimes \mathcal{H}_6))$ be two quantum channels, and $M \equiv M_{3120} \in \mathcal{L}(\mathcal{H}_3 \otimes \mathcal{H}_1 \otimes \mathcal{H}_2\otimes \mathcal{H}_0$ and $N \equiv N_{6435} \in \mathcal{L}(\mathcal{H}_6 \otimes \mathcal{H}_4 \otimes \mathcal{H}_3 \otimes \mathcal{H}_5)$ be the corresponding Choi operators of $\mathcal{M}$ and $\mathcal{N}$, respectively. We define $V_1:
	\mathcal{H}_0 \otimes \mathcal{H}_2\longrightarrow \mathcal{H}_1 \otimes \mathcal{H}_3\otimes \mathcal{H}_{A_1},V_2:\mathcal{H}_3 \otimes \mathcal{H}_5 \longrightarrow  \mathcal{H}_4 \otimes \mathcal{H}_6 \otimes \mathcal{H}_{A_2}$ as
	\begin{align}
	V_1&=(I_{13} \otimes (M^{T})^{1/2})(|I_{13}\rangle\rangle \otimes I_{02}),\\
	V_2&=(I_{46}\otimes(N^{T})^{1/2})(|I_{46}\rangle\rangle \otimes I_{35})
	\end{align}
	and $$V=(I_1\otimes V_2\otimes I_{A_1})\circ (V_1\otimes I_5),\mathcal{H}_A =\mathcal{H}_{A_1}\otimes\mathcal{H}_{A_2}.$$ Then we have
	\begin{equation}
	\mathcal{N} \star \mathcal{M}(\rho)=(\mathcal{I}_1 \otimes \mathcal{N})\circ (\mathcal{M}\otimes \mathcal{I}_5)(\rho)=\textup{Tr}_{A_1A_2}[V\rho V^{\dagger}],
	\end{equation}
	where $I_{A_1}$ is the identity map on $\mathcal{H}_{A_1}$, $I_{1}$ is the identity map on $\mathcal{H}_{1}$ and $V^{\dagger}V=I_{025}$.
\end{thm}

\begin{thm}
	Let $\mathcal{M} \in \mathcal{L}(\mathcal{L}(\mathcal{H}_0 \otimes \mathcal{H}_2),\mathcal{L}(\mathcal{H}_1 \otimes \mathcal{H}_3))$ and $\mathcal{N} \in \mathcal{L}(\mathcal{L}(\mathcal{H}_3 \otimes \mathcal{H}_5),\mathcal{L}(\mathcal{H}_4 \otimes \mathcal{H}_6))$ be two quantum channels. We have $\mathcal{N} \star \mathcal{M} :=(\mathcal{I}_1 \otimes \mathcal{N}) \circ (\mathcal{M} \otimes \mathcal{I}_5)$, where $(\mathcal{M}\otimes\mathcal{I}_5) \in \mathcal{L}(\mathcal{L}(\mathcal{H}_0 \otimes \mathcal{H}_2 \otimes \mathcal{H}_5),\mathcal{L}(\mathcal{H}_1 \otimes \mathcal{H}_3 \otimes \mathcal{H}_5))$ and $(\mathcal{I}_1 \otimes \mathcal{N}) \in \mathcal{L}(\mathcal{L}(\mathcal{H}_1 \otimes \mathcal{H}_3 \otimes \mathcal{H}_5),\mathcal{L}(\mathcal{H}_1 \otimes \mathcal{H}_4 \otimes \mathcal{H}_6))$. $M\in\mathcal{L}(\mathcal{H}_1 \otimes \mathcal{H}_3\otimes\mathcal{H}_0 \otimes \mathcal{H}_2)$ and $N\in\mathcal{L}(\mathcal{H}_4 \otimes \mathcal{H}_6\otimes\mathcal{H}_3 \otimes \mathcal{H}_5)$ be their corresponding Choi operators. Then we have $\mathcal{M} \in \mathcal{L}(\mathcal{L}(\mathcal{H}_0 \otimes \mathcal{H}_2 \otimes \mathcal{H}_5),\mathcal{L}(\mathcal{H}_1 \otimes \mathcal{H}_3 \otimes \mathcal{H}_5)),\mathcal{N} \in \mathcal{L}(\mathcal{L}(\mathcal{H}_1 \otimes \mathcal{H}_3 \otimes \mathcal{H}_5),\mathcal{L}(\mathcal{H}_1 \otimes \mathcal{H}_4 \otimes \mathcal{H}_6))$, $\mathcal{N} \star \mathcal{M} :=(\mathcal{I}_1 \otimes \mathcal{N}) \circ (\mathcal{M} \otimes \mathcal{I}_5)$, and the Choi
	operator of $\mathcal{N}\star\mathcal{M}$ is $N\ast M=\text{Tr}[(I_{456}\otimes M^{T_3})(N\otimes I_{012})]$. Define $V:\mathcal{H}_0\otimes\mathcal{H}_2\otimes\mathcal{H}_5\longrightarrow \mathcal{H}_1\otimes\mathcal{H}_4\otimes\mathcal{H}_6\otimes\mathcal{H}_A$ as
	\begin{equation}
	V=(I_{146}\otimes(N\ast M)^{T/2})(|I_{146}\rangle\rangle\otimes I_{025}).
	\end{equation}
	such that
	\begin{equation}
	V^{\dagger}V=I_{025},
	\end{equation}
	\begin{equation}
	\mathcal{N}\star\mathcal{M}(\rho)= \textup{Tr}[V\rho V^{\dagger}].
	\end{equation}
\end{thm}

\section{Discrimination of quantum channels}
Given two density operators $\rho$ and $\sigma$, the fidelity
$$F(\rho, \sigma)=\text{Tr}\sqrt{\sqrt{\rho}\sigma\sqrt{\rho}}$$
quantifies the extent to which $\rho$ and $\sigma$ can be distinguished from one another \cite{fidelity}. It is obvious that $0 \leq F(\rho,\sigma)\leq 1$, $F(\rho,\sigma)=1$ if and
only if $\rho\ = \sigma$ ($\rho$ and $\sigma$ are indistinguishable) and $F(\rho,\sigma)=0$ if and
only if $\rho\ \perp \sigma$ ($\rho$ and $\sigma$ are completely distinguishable).

For quantum channels, the fidelity is defined by the Choi operators. Let $\mathcal{M},\mathcal{N}\in \mathcal{L}(\mathcal{L}(\mathcal{H}_0),\mathcal{L}(\mathcal{H}_1))$ and $M,N\in \mathcal{L}(\mathcal{H}_1 \otimes \mathcal{H}_0)$ be their corresponding Choi operators. The fidelity of $\mathcal{M}$ and $\mathcal{N}$ is defined by
\begin{align}
\mathcal{F}(\mathcal{M}, \mathcal{N})=F\big(\frac{M}{d_0},\frac{N}{d_0} \big),
\end{align}
where $d_0 =\textup{dim}(\mathcal{H}_0)$ \cite{fidelity}. Similar to the fidelity of density operators, when $\mathcal{M}$ and $\mathcal{N}$ are orthonormal, that is $\mathcal{F}(\mathcal{M},\ \mathcal{N})=0$, $\mathcal{M}$ and $\mathcal{N}$ can be completely distinguished; while when $\mathcal{M}$ is not orthogonal to $\mathcal{N}$, then $0 < \mathcal{F}(\mathcal{M},\mathcal{N}) < 1$,
$\mathcal{M}$ and $\mathcal{N} $ cannot be fully distinguished.

Now we show that the distinguishability of quantum channels $\mathcal{M}$ and $\mathcal{N}$ can be improved by considering link product of each quantum channel $n$ times, as $n$ grows.
For that we first take into account the following two lemmas.

\begin{lem}
	Let $\mathcal{M},\mathcal{N}\in \mathcal{L}(\mathcal{L}(\mathcal{H}_0),\mathcal{L}(\mathcal{H}_1))$ be two quantum channels and $M,N$ be their corresponding Choi operators. Then $M \otimes N$ is the Choi operator of $\mathcal{M} \otimes \mathcal{N}$.
\end{lem}
\begin{proof}
	Let $T$ be the Choi operator of $\mathcal{M} \otimes \mathcal{N}$. According to the Choi operator definition, we have
	\begin{align}\label{zhangliang}
	T&=\mathcal{M} \otimes \mathcal{N} \otimes \mathcal{I}_{00}(|I_{00}\rangle\rangle\langle\langle I_{00})|\notag\\
	&=((\mathcal{M} \otimes \mathcal{I}_0)(|I_0\rangle\rangle\langle\langle I_{0}|))\otimes ((\mathcal{N} \otimes \mathcal{I}_0)(|I_0\rangle\rangle\langle\langle I_{0}|))\notag\\
	&=M \otimes N .
	\end{align}
\end{proof}

\begin{lem}\label{mm}
	Let $\mathcal{M}\in \mathcal{L}(\mathcal{L}(\mathcal{H}_0),\mathcal{L}(\mathcal{H}_1))$ be a quantum channel and $M$ be its corresponding Choi operator. Then $\mathcal{M} \star \mathcal{M}=\mathcal{M}^{\otimes2}$.
\end{lem}
\begin{proof}
	We have $\mathcal{M} \star \mathcal{M}=(\mathcal{M} \otimes \mathcal{I}_1)\circ(\mathcal{I}_0 \otimes \mathcal{M})$,
	where $(\mathcal{M} \otimes \mathcal{I}_1) \in \mathcal{L}(\mathcal{L}(\mathcal{H}_0 \otimes \mathcal{H}_1),\mathcal{L}(\mathcal{H}_1 \otimes \mathcal{H}_1)),\ (\mathcal{I}_0 \otimes \mathcal{M}) \in \mathcal{L}(\mathcal{L}(\mathcal{H}_0 \otimes \mathcal{H}_0),\mathcal{L}(\mathcal{H}_0 \otimes \mathcal{H}_1))$ and $\mathcal{M} \star \mathcal{M}\in \mathcal{L}(\mathcal{L}(\mathcal{H}_0^{\otimes 2}),\mathcal{L}(\mathcal{H}_1^{\otimes 2}))$. 
And	$\mathfrak{C}(\mathcal{M}\star \mathcal{M})=(\mathcal{M} \otimes \mathcal{I}_1)(\mathcal{I}_0 \otimes \mathcal{M})(|\Phi^{+}\rangle\langle\Phi^{+}|)$, where $|\Phi^+\rangle\in\mathcal{H}_0^{\otimes 2}\otimes\mathcal{H}_0^{\otimes 2}$ is a maximally entangled state given by $|\Phi^{+}\rangle = \sum_{i=1}^{d_0}|e_i\rangle|e_i\rangle\otimes|e_i\rangle|e_i\rangle$, with $\{|e_i\rangle\}_{i=1}^{d_0}$ being an orthonormal basis for $\mathcal{H}_0$.
	
	The Choi operator of $\mathcal{M}\star\mathcal{M}$ is given as
	$$\begin{aligned}
	\mathfrak{C}(\mathcal{M}\star \mathcal{M})&=(\mathcal{M} \otimes \mathcal{I}_1)(\mathcal{I}_0 \otimes \mathcal{M})(|\Phi^{+}\rangle\langle\Phi^{+}|)\\
	&=(\mathcal{M} \otimes \mathcal{I}_1)(|\Psi^{+}\rangle\langle\Psi^{+}|\otimes M)\\
	&=M \otimes M,
	\end{aligned}$$
	where $|\Psi^{+}\rangle=\sum_{i=1}^{d_0}|e_i\rangle|e_i\rangle$.
	
	Also, by Lemma 3.1, the Choi operator of $\mathcal{M}^{\otimes 2}$ is $M\otimes M$. As the Choi representation is isomorphic, we immediately have $$\mathcal{M}\star\mathcal{M}=\mathcal{M}^{\otimes 2}.$$
\end{proof}
\begin{rmk}
	The above is an obvious conclusion for finite dimension but it is not necessarily true for infinite dimension. Because in infinite dimension $(A\otimes C)(B\otimes D)$ is not necessarily equal to $AB\otimes CD$. But for $\mathcal{M}\in \mathcal{L}(\mathcal{L}(\mathcal{H}_0),\mathcal{L}(\mathcal{H}_1))$ where $\textup{dim}(\mathcal{H}_0)=\infty,\textup{dim}(\mathcal{H}_1)=\infty$ and $\mathcal{M}$ is a compact operator, we can prove that the above result is correct. That is, 
$\mathcal{M}^{\star n}=\underbrace{\mathcal{M}\star\mathcal{M}\star\cdots\star\mathcal{M} }_n = \mathcal{M}^{\otimes n}$.
\end{rmk}

\begin{thm} (Distinguishability of quantum channels)
	Let $\mathcal{M},\ \mathcal{N}\in \mathcal{L}(\mathcal{L}(\mathcal{H}_0),\ \mathcal{L}(\mathcal{H}_1))$ be two quantum channels. Suppose $\mathcal{M}\neq \mathcal{N}.$
	Then  $$\lim_{n \rightarrow \infty }\mathcal{F}(\mathcal{M}^{\star n} ,\mathcal{N}^{\star n})=0.$$
\end{thm}

\begin{proof}
	According to Lemma \ref{mm}, we have
	$$\mathcal{F}(\mathcal{M}^{\star n},\ \mathcal{N}^{\star n})=\mathcal{F}(\mathcal{M}^{\otimes n},\ \mathcal{N}^{\otimes n})=\mathcal{F}(\mathcal{M},\ \mathcal{N})^n.$$ Because $0 < \mathcal{F}(\mathcal{M},\ \mathcal{N}) < 1$, we have
	\begin{align}
	1 > \mathcal{F}(\mathcal{M},\ \mathcal{N}) > \mathcal{F}(\mathcal{M},\mathcal{N})^{2} > \ldots <\mathcal{F}(\mathcal{M}, \mathcal{N})^{n}  > 0.
	\end{align}
	Thus $\mathcal{F}(\mathcal{M}^{\star n} ,\mathcal{N}^{\star n})$ decreases gradually and strictly. Therefore, we have $$\lim_{n \rightarrow \infty }\mathcal{F}(\mathcal{M}^{\star n} ,\mathcal{N}^{\star n})= \lim_{n \rightarrow \infty}\mathcal{F}(\mathcal{ M}, \mathcal{N})^n=0.$$ For $\forall \varepsilon >0$, $\exists n$ (a natural number) and $ \tilde{N}=[log_{\mathcal{F}(\mathcal{M},\mathcal{N})}\varepsilon]$, such that when $n> \tilde{N}$,  $0< \mathcal{F}(\mathcal{M}^{\star n} ,\mathcal{N}^{\star n}) <\varepsilon$.
	Thus the distinguishability of $\mathcal{M}$ and $\mathcal{N}$ is improved gradually as $n$ grows.
\end{proof}

\section{Uhlmann's theorem for the fidelity of diagonal channels}
Let $\mathcal{M}\in\mathcal{L}(\mathcal{L}(\mathcal{H}_0),\mathcal{L}(\mathcal{H}_1))$ be a quantum channel. Suppose  $\text{dim}(\mathcal{H}_0)=\text{dim}(\mathcal{H}_1)=2$.
We consider an orthonormal basis of the Hermitian matrix \cite{2D,diag},\\
$$\beta=\{\frac{\sigma_0}{\sqrt{2}}, \frac{\sigma_1}{\sqrt{2}},\frac{\sigma_2}{\sqrt{2}},\frac{\sigma_3}{\sqrt{2}}\},$$
where $\sigma_i$ are the Pauli matrices,
$$\sigma_0=I,
\sigma_1=\begin{pmatrix}
0 & 1 \\
1 & 0
\end{pmatrix},
\sigma_2=\begin{pmatrix}
0 & -i \\
i & 0
\end{pmatrix},
\sigma_3=\begin{pmatrix}
1 & 0 \\
0 & -1
\end{pmatrix}.$$
If the representation of the quantum channel $\mathcal{M}$ with respect to the basis $\beta$ is diagonal, i.e.,$$\mathcal{M}=\textup{diag}(1,a_1,a_2,a_3),$$
then $\mathcal{M}$ is called a diagonal channel \cite{diag}.
There are four important families of two-dimensional diagonal channels:
\begin{align*}
&1.\  \mathcal{C}=\textup{diag}(1,p,p,p),\ \text{where}\ -\frac{1}{3}\leq p\leq 1;\\
&2.\  \mathcal{D}=\textup{diag}(1,p,-p,p),\ \text{where}\  -1 \leq p\leq \frac{1}{3};\\
&3.\  \mathcal{R}=\textup{diag}(1,-p,-p,p),\ \text{where}\ -\frac{1}{3}\leq p\leq 1;\\
&4.\  \mathcal{S}=\textup{diag}(1,p,-p,p),\ \text{where}\ -1 \leq p\leq \frac{1}{3}.\\
\end{align*}
Quantum channels $\mathcal{C}, \mathcal{D}, \mathcal{R}$, and $\mathcal{S}$ are called depolarizing, transpose depolarizing,
hybrid depolarizing classical, and hybrid transpose depolarizing classical, respectively.
They are among the most widely used channels in science and technology \cite{2D,2003capacity,2006remark,2006additivity,2009unital}.

Let  $C,D,R$ and $S$ be the Choi operators of $\mathcal{C}, \mathcal{D}, \mathcal{R}$ and $\mathcal{S}$, respectively. According to the calculation method of Choi operator in \cite{diag}, we can obtain
\begin{align*}
C&=\begin{pmatrix}
\frac{1+p}{2} & 0 & 0 & p \\
0 & \frac{1-p}{2} & 0 & 0 \\
0 & 0 & \frac{1-p}{2} & 0 \\
p & 0 & 0 & \frac{1+p}{2}
\end{pmatrix},
D=\begin{pmatrix}
\frac{1+p}{2} & 0 & 0 & 0 \\
0 & \frac{1-p}{2} & P & 0 \\
0 & P & \frac{1-p}{2} & 0 \\
0 & 0 & 0 & \frac{1+p}{2}
\end{pmatrix}, \\
R&=\begin{pmatrix}
\frac{1+p}{2} & 0 & 0 & -p \\
0 & \frac{1-p}{2} & 0 & 0 \\
0 & 0 & \frac{1-p}{2} & 0 \\
-p & 0 & 0 & \frac{1+p}{2}
\end{pmatrix},
S=\begin{pmatrix}
\frac{1+p}{2} & 0 & 0 & 0 \\
0 & \frac{1-p}{2} & -P & 0 \\
0 & -P & \frac{1-p}{2} & 0 \\
0 & 0 & 0 & \frac{1+p}{2}
\end{pmatrix}.
\end{align*}

Also, we have
\[
\sqrt{C} =
\begin{pmatrix}
\frac{\sqrt{1-p}}{2\sqrt{2}}+\frac{\sqrt{3p+1}}{2\sqrt{2}} & 0 & 0 & -\frac{\sqrt{1-p}}{2\sqrt{2}}+\frac{\sqrt{3p+1}}{2\sqrt{2}} \\
0 & \sqrt{\frac{1-p}{2}} & 0 & 0 \\
0 & 0 & \sqrt{\frac{1-p}{2}} & 0 \\
-\frac{\sqrt{1-p}}{2\sqrt{2}}+\frac{\sqrt{3p+1}}{2\sqrt{2}} &  0 & 0 & \frac{\sqrt{1-p}}{2\sqrt{2}}+\frac{\sqrt{3p+1}}{2\sqrt{2}}
\end{pmatrix},
\]

and 
\begin{align*}
\sqrt{C}D\sqrt{C} =\begin{pmatrix}
\frac{(p+1)^2}{4} & 0 & 0 & \frac{p(p+1)}{2}\\
0 & \frac{(1-p)^2}{4} & \frac{p(1-p)}{2} & 0\\
0 &  \frac{p(1-p)}{2} & \frac{(1-p)^2}{4} & 0\\
\frac{p(p+1)}{2} & 0 & 0 & \frac{(p+1)^2}{4}\\
\end{pmatrix}.
\end{align*}

We note that $\sqrt{C}D\sqrt{C}=CD=\sqrt{C}\sqrt{C}D.$
When $C$ is reversible, it is obvious that $D\sqrt{C}=\sqrt{C}D$. When $C$ is irreversible, since $C\geq 0$, for $\forall \varepsilon >0$ we have $\sqrt{C}+\varepsilon I>0$ and
\begin{align}
(\sqrt{C}+\varepsilon I)D(\sqrt{C}+\varepsilon I)=\sqrt{C}D\sqrt{C}+\varepsilon \sqrt{C}D+\varepsilon D\sqrt{C}+\varepsilon D,\notag\\
(\sqrt{C}+\varepsilon I)(\sqrt{C}+\varepsilon I)D=\sqrt{C}\sqrt{C}D+\varepsilon \sqrt{C}D+\varepsilon D\sqrt{C}+\varepsilon D,\notag
\end{align}
so that $$(\sqrt{C}+\varepsilon I)D(\sqrt{C}+\varepsilon I)=(\sqrt{C}+\varepsilon I)(\sqrt{C}+\varepsilon I)D.$$
On the left multiplication by $(\sqrt{C}+\varepsilon I)^{-1}$, we have $$D(\sqrt{C}+\varepsilon I)=(\sqrt{C}+\varepsilon I)D,$$
$$D\sqrt{C}=\sqrt{C}D.$$
This proves that $\sqrt{C}$ and hence $C$ can be interchanged with $D$. 
%
Similarly, we find that $MN=NM$, where $M, N \in \{C,D,R,S\}$. That is $C,D,R$ and $S$ are all interchangeable.

Using the above results, we can get the following theorem easily.
\begin{thm}
	For any two channels $\mathcal{M},\mathcal{N} \in \{\mathcal{C},\mathcal{D},\mathcal{R},\mathcal{S}\}$,
	\begin{align}
	\mathcal{F}(\mathcal{M},\mathcal{N})=\frac{1}{2}\sum_{i=1}^4 \sqrt{\alpha_i\beta_i},
	\end{align}
	where $\alpha_i, \beta_i, i=1, 2, 3, 4$ are the eigenvalues of $M,N$ respectively and $M,N$ are Choi operators of $\mathcal{M},\mathcal{N}$.
\end{thm}

Uhlmann's theorem for the fidelity of quantum channels \cite{fidelity} is stated as
\begin{equation}
\mathcal{F}(\mathcal{M},\mathcal{N})=\frac{1}{2}\max_{V,W}|\textup{Tr}(V^{\dagger} W)|,
\end{equation}
where $V$ and $W$ are isometries $\mathcal{H}_0 \longrightarrow \mathcal{H}_1 \otimes \mathcal{H}_A$ that define Stinespring dilations of $\mathcal{M}$ and $\mathcal{N}$.\\

For the diagonal channels, we find that the maximum value of Uhlmann's theorem is achievable.\
\begin{thm}
	Let $\mathcal{M},\mathcal{N}\in\{\mathcal{C},\mathcal{D},\mathcal{R},\mathcal{S}\}$. There exist $V_0$ and $W_0$ such that
	\begin{align}
	\mathcal{F}(\mathcal{M}^{\star n},\mathcal{N}^{\star n})=\frac{1}{2^n}|\textup{Tr}(V_0^{\dagger} W_0)|^n,
	\end{align}
	where $V_0$ and $W_0$ are isometries that define Stinespring dilations of $\mathcal{M}$ and $\mathcal{N}$ respectively.
\end{thm}
\begin{proof}
	Let $V_0,W_0:\mathcal{H}_0 \longrightarrow \mathcal{H}_1 \otimes \mathcal{H}_A$ as
	\begin{align}
	V_0=(I_1\otimes (M^T)^{\frac{1}{2}})(|I_1\rangle\rangle\otimes I_0),\\
	W_0=(I_1\otimes (N^T)^{\frac{1}{2}})(|I_1\rangle\rangle\otimes I_0),
	\end{align}
	where $M$ and $N$ are the Choi operators of $\mathcal{M}$ and $\mathcal{N}$ respectively, and $\mathcal{H}_A=\textup{Supp}(M)\ \textup{or}\ \textup{Supp}(N)$.
	Thus, we have
	\begin{align*}
	V_0^{\dagger}W_0 &=(I_1\otimes (M^T)^{\frac{1}{2}})(|I_1\rangle\rangle\otimes I_0)^{\dagger}(I_1\otimes (N^T)^{\frac{1}{2}})(|I_1\rangle\rangle\otimes I_0)\\
	&=(\langle\langle I_1|\otimes I_0)(I_1\otimes (M^T)^{\frac{1}{2}})(|I_1\rangle\rangle\otimes I_0)^{\dagger}(I_1\otimes (N^T)^{\frac{1}{2}})\\
	&=\textup{Tr}_1[(M^TN^T)^{\frac{1}{2}}],
	\end{align*}
	and
	\begin{align*}
	\frac{1}{2}\textup{Tr}(V_0^{\dagger} W_0)=\frac{1}{2}\textup{Tr}[(M^TN^T)^{\frac{1}{2}}]=\frac{1}{2}\textup{Tr}[(MN)^{\frac{1}{2}}]=\mathcal{F}(\mathcal{M},\mathcal{N}).
	\end{align*}
	Thus, for any $\mathcal{M},\mathcal{N} \in \{\mathcal{C},\mathcal{D},\mathcal{R},\mathcal{S}\}$, we have
	\begin{equation}
	\mathcal{F}(\mathcal{M},\mathcal{N})=\frac{1}{2}|\textup{Tr}(V_0^{\dagger} W_0)|
	\end{equation}
	and
	\begin{equation}
	\mathcal{F}(\mathcal{M}^{\star n},\mathcal{N}^{\star n})=\mathcal{F}(\mathcal{M},\mathcal{N})^n=\frac{1}{2^n}|\textup{Tr}(V_0^{\dagger} W_0)|^n \xrightarrow{n \rightarrow \infty} 0.
	\end{equation}
\end{proof}

\section{Conclusion} In summary, we have established the Stinespring dilation theorem of the link product of quantum channels in two different ways and discussed the discrimination of quantum channels. We have shown that the distinguishability can be improved by self-linking each quantum channel $n$ times as $n$ grows. We also found that the maximum value of Uhlmann's theorem can be achieved for diagonal channels.

\vspace{3cm}

\thanks{{\bf Acknowledgement.} This  project is supported by National Natural Science Foundation of China (Grants No. 61877054, 12031004, and 12271474).

\bibliographystyle{amsplain}

\end{document}